 \documentstyle[preprint,aps]{revtex} 

\input epsf

\begin{document}

\draft 

\tighten 
\preprint{\vbox{\hbox{CLNS 94/1274 \hfill}
                \hbox{CLEO 94--9  \hfill}
                \hbox{\today       \hfill}}}

\title{Two-Photon Production of Charged Pion and Kaon Pairs}

\author{
J.~Dominick,$^{1}$ M.~Lambrecht,$^{1}$ S.~Sanghera,$^{1}$
V.~Shelkov,$^{1}$ T.~Skwarnicki,$^{1}$ R.~Stroynowski,$^{1}$
I.~Volobouev,$^{1}$ G.~Wei,$^{1}$ P.~Zadorozhny,$^{1}$
M.~Artuso,$^{2}$
% ...
}

\address{
{\rm (CLEO Collaboration)}\\  
$^{1}${Southern Methodist University, Dallas, Texas 75275}\\
$^{2}${Syracuse University, Syracuse, New York 13244}\\
% ...
}        

\date{\today}
\maketitle


\begin{abstract}
A measurement of the cross section for the combined two-photon
production of charged pion and kaon
pairs is performed using 1.2~$\rm fb^{-1}$ of data collected by
the CLEO~II detector at the Cornell Electron Storage Ring.
The cross section is measured at invariant masses of the two-photon
system between 1.5 and 5.0~GeV/$c^2$, and at scattering angles more
than $53^\circ$ away from the $\gamma\gamma$ collision axis in the
$\gamma\gamma$ center-of-mass frame.
The large background of leptonic
events is suppressed by utilizing the CsI
calorimeter in conjunction with the muon chamber system.
The reported cross section is compared with leading order QCD models
as well as previous experiments.  In particular,
although the functional dependence of the measured cross section
disagrees with leading order QCD at small values of the two-photon
invariant mass, the data show a transition to
perturbative behavior at an invariant mass of approximately 2.5~GeV/$c^2$.
\end{abstract}
%


\section{\label{sec:intro} Introduction}

A measurement of the continuum production of meson pairs in two-photon
reactions provides a test of quantum chromodynamics
(QCD) for exclusive
processes\cite{ref:Brodsky,ref:Gunion,ref:Benayoun,ref:Nizic,ref:Ji}.
The leading order result based on perturbative QCD
has been calculated by Brodsky and Lepage\cite{ref:Brodsky} for large
center-of-mass scattering angle $\theta^*$
and large invariant mass
$W$ of the two-photon system.  Extensions of this work to finite
values of $q^2$ (the invariant mass of the virtual photon) for one or
both of the photons, as well as
the incorporation of SU(3) flavor symmetry breaking effects are
discussed
by Gunion, Miller, and Sparks\cite{ref:Gunion}, and Benayoun and
Chernyak\cite{ref:Benayoun} respectively.
A calculation of the
next-to-leading-order QCD prediction based on a particular model of
the meson distribution amplitude is discussed
by Ni{\v z}i{\' c}\cite{ref:Nizic},
while the running of $\alpha_{\rm s}$ in the calculation of the meson form
factor, and the subsequent effect on the leading order QCD result for
$\gamma\gamma\rightarrow M^+M^-$ is discussed
by Ji and Amiri\cite{ref:Ji}.

The subject of this article is a measurement of the two-photon production of
$\pi^+\pi^-$ and $K^+K^-$
using the CLEO~II detector
located at the Cornell Electron Storage Ring (CESR).
The leading order result
calculated by Brodsky and Lepage\cite{ref:Brodsky} can be expressed,
approximately, in terms of
the cross section for the two-photon production of muon pairs:
\begin{equation}
{{d\sigma}\over{d\cos\theta^*}}\; (\gamma\gamma\rightarrow M^+M^-) \approx
{{4\;|F_{\rm M}(W^2)|^2}\over{1-\cos^4\theta^*}}\;\;
{{d\sigma}\over{d\cos\theta^*}}\; (\gamma\gamma\rightarrow \mu^+\mu^-)
\label{eq:Brodsky}
\end{equation}
The angle $\theta^*$ denotes the scattering angle of one of the
charged particles (``prongs'') with respect to the $\gamma\gamma$ collision
axis, calculated in the $\gamma\gamma$ center-of-mass frame.
The overall angular dependence of Eq.~(\ref{eq:Brodsky})
is $\sin^{-4}\theta^*$ when one folds in the muon pair cross section.
The form factor $F_{\rm M}$ depends on the strong coupling constant
$\alpha_{\rm s}$, the $\gamma\gamma$ invariant mass $W$, and the
pseudoscalar decay constant $f_{\rm M}$.  It is
taken to be $(0.4~{\rm GeV}^2/c^4)/W^2$ for pions\cite{ref:Brodsky}.
A residual dependence on the pseudoscalar
wavefunction outside of the form factor dependence is ignored by
Eq.~(\ref{eq:Brodsky}), and leads to a $\sim 25\%$ negative
correction.  This model dependent correction is significantly
larger for the neutral mesons. Since the form factor
is proportional to $f_{\rm M}^2$, the ratio of the kaon pair cross section to
the pion pair cross section depends on $(f_{\rm K}/f_\pi)^4 =
2.2$\cite{ref:PDG}
when differences in the $\pi$ and $K$ wavefunctions are ignored.
Benayoun and Chernyak\cite{ref:Benayoun} argue that such SU(3) flavor symmetry
breaking effects are important and obtain a ratio closer to unity, but
their absolute prediction for the combined pion pair plus kaon pair
cross section does not differ appreciably from that of Brodsky and Lepage.
The insensitivity of the combined cross section for charged pion
and kaon pairs
to the choice of wavefunction models makes this measurement an ideal probe
of leading order perturbative QCD.

Both the Mark~II and the TPC/Two-Gamma collaborations have
published results\cite{ref:MarkII,ref:TPCpipi} on the high mass
two-photon continuum production of charged pion and kaon pairs using the PEP
$e^+e^-$ storage ring at SLAC.
Mark~II data on the combined $\pi^+\pi^-$ and
$K^+K^-$ cross section, measured in the dipion mass range between 1.7
and 3.5~GeV/$c^2$, suggest agreement with
leading order QCD at values of $W_{\pi\pi}$ larger than 2~GeV/$c^2$
($|\cos\theta^*|<0.5$).
TPC/Two-Gamma data on the $K^+K^-$
production cross section show reasonable agreement with the leading order QCD
prediction in the measured $W$ range from 1.3 to 3.5~GeV/$c^2$ for
$|\cos\theta^*|<0.6$. Charged pion pair data are about 3 times larger than the
leading order QCD prediction for $W$ measured from 1.5 to 2.5~GeV/$c^2$ and
$|\cos\theta^*|<0.3$, but at least some of the excess is likely to be
caused by the interference of the $f_2(1270)$ resonance with the
$\pi^+\pi^-$ continuum.
The leading order perturbative QCD calculations become more reliable
at large $W$, so the experimental goal of a measurement of these
two-photon processes is to probe larger values of $W$ than previously
obtained using higher statistics to observe the transition in the cross
section from non-perturbative to perturbative behavior.

We report on a measurement of the
combined cross section for charged pion and kaon pairs using the
CLEO~II detector at CESR.
The measurement takes advantage of a much larger data sample than
available at previous experiments, as well as a larger angular
acceptance ($|\cos\theta^*|<0.6$).  The cross section is reported in
the $W_{\pi\pi}$ range from 1.5 to 5.0~GeV/$c^2$ using 1.23~$\rm fb^{-1}$
of integrated
luminosity at a center-of-mass energy on or near the $\Upsilon(4S)$
resonance ($\sqrt{s}\approx 10.6$~GeV).
A related study on the two-photon production of proton-antiproton
pairs has recently been reported\cite{ref:CLEOppbar}.

\section{\label{sec:setup} Experimental Setup}
\subsection{Detector}
The CLEO~II detector\cite{ref:CLEOnim} is a general purpose solenoidal detector
designed to make precision measurements of both charged and neutral
particles.  Charged particle tracking is accomplished through the use
of a 51 layer proportional drift chamber accompanied by a 10 layer
intermediate wire chamber and a 6 layer straw tube chamber located
just outside of the beryllium beam pipe.
The transverse momentum resolution achieved within the 1.5~T field provided by
the solenoidal magnet is given by $(\delta p_\perp/p_\perp)^2 =
(0.11\%p_\perp)^2 + (0.67\%)^2$, with $p_\perp$ expressed in GeV/$c$.
The tracking system is
surrounded by both a time-of-flight scintillation system, consisting
of 64 counters arranged in $\phi$ (the azimuthal coordinate transverse
to the beam axis) plus 28 counters per endcap, and an
electromagnetic calorimeter composed of 7800 Thallium-doped Cesium
Iodide (CsI) crystals.  The barrel section of the CsI calorimeter
forms an integral part of this analysis as the fine tower segmentation
allows for particle discrimination based on an analysis of shower shapes.
The projective crystals are approximately 5~cm square at the front
end and 30~cm long.  The measured electromagnetic energy resolution
is given by $\delta E/E = 0.35\%/E^{0.75} + 1.9\% - 0.1\% E$, with
$E$ expressed in GeV.
All of these detector components lie within the
1.5~T superconducting solenoidal coil, which is in turn surrounded
by 3 layers of iron interleaved with 3 sets of tracking chambers for
muon detection.  Each tracking chamber set contains three planar
layers of Iarocci tubes with cathode strips providing information on
the coordinate orthogonal to the wires.  The barrel muon system is
octagonal in cross section and the angular acceptance of the
innermost superlayer is $|\cos \theta| < 0.71$ ($\theta$ is
polar angle measured with respect to the beam axis).

\subsection{\label{sec:trigger} Trigger}
The trigger forms one of the more challenging aspects
of an analysis of low multiplicity final states since it must
meet the conflicting demands of high efficiency yet large
discrimination against beam-wall and beam-gas backgrounds.  The
components of the CLEO~II detector used for this purpose include the
tracking chambers, time-of-flight scintillation counters, and
calorimeter crystals---all of which are combined into a three level
trigger system.\cite{ref:CLEOtrig}

The two-prong trigger used in this analysis
is designed to collect events containing either two minimum ionizing
particles or at least one showering particle (electron or photon).
The requirements of this trigger are the following:
First, the time-of-flight system must register hits in at least two
non-adjacent barrel scintillation counters.  Second, the CsI calorimeter
must identify either two separated energy clusters consistent with
minimum ionizing particles ($\sim0.2$~GeV of deposited energy), or
one energy cluster consistent with a showering particle
($>0.5$~GeV of deposited energy).  Third, the tracking chambers must
exhibit two clean track patterns such that the momentum transverse to
the beam direction ($p_\perp$) of one track is at least 0.4~GeV/$c$,
while that of the other is at least 0.2~GeV/$c$.  The exact trigger
specification depends on the running period of the experiment, but at
least one of the tracks must extend radially through the entire drift
chamber.

Several methods were devised to measure the two-prong trigger
efficiency from data.  One technique makes use of events that satisfy
a looser (prescaled) trigger to measure the efficiency of the
additional trigger conditions in the standard two-prong trigger.
Another technique exploits the redundancy of a calorimeter-only
trigger to measure the efficiency of the tracking components of the
two-prong trigger. Finally, a third technique uses four-prong events
where two tracks are sufficient to fire the two-prong trigger. The
trigger efficiency of the remaining two tracks is then studied.
All three techniques are in very good agreement on the efficiency of
those trigger components that are in common.  The
measured trigger efficiencies are listed separately for each trigger
system in Table~\ref{tab:trigger}. The measurements
refer to the average efficiency for two-prong
events in the fiducial region $|\cos\theta|<0.7$ and with each track
satisfying $p_\perp>0.5$~GeV/$c$.  When all trigger conditions are applied,
the overall trigger efficiency for
charged pion and kaon pairs is estimated to be 79\% or 86\% depending upon
the running period of the experiment.  A Monte Carlo simulation of
the trigger system, calibrated to an independent data sample,
confirms these values to within 5\%. Therefore, the systematic uncertainty
on the trigger efficiency is estimated to be 5\%.

The two-prong trigger runs at a rate of
approximately 7~Hz at a luminosity of $2\times 10^{32}{\rm cm^{-2}s^{-1}}$.
All of the events collected by the CLEO~II
trigger are fed into a software filter to remove obvious beam-gas and
beam-wall backgrounds.  This filter initially ran offline for the
first 500~$\rm pb^{-1}$ of data used in this analysis, but presently
runs online as a part of the trigger system.  It makes extensive
use of all the tracking devices to reject events without clean track
patterns originating from the nominal interaction point.  Its efficiency can be
measured directly from data since a prescaled sample of rejected events is
saved for analysis, but the efficiency can also be determined by applying the
filter to Monte Carlo events.  For two-photon two-prong events, this efficiency
is roughly 95\%.

\section{\label{sec:analysis} Data Analysis}

In this section we address the details of the analysis techniques
used to extract the cross section for the two-photon production of charged
pion and kaon pairs. First, the details of the event generation
and detector simulation are discussed in Sec.~\ref{sec:mc}.
Second, the event shape cuts used to select two-photon events from the data
sample are described in Sec.~\ref{sec:selection}; and third, a technique
for suppressing the large muon background in the two-photon event sample
using information from the CsI calorimeter is presented
in Sec.~\ref{sec:rejection}.

\subsection{\label{sec:mc} Event Simulation}
The extraction of the measured cross section for
$\gamma\gamma\rightarrow M^+M^-$ ($M=\pi, K$)
relies on an accurate calculation of
the QED process $e^+e^-\rightarrow e^+e^-\gamma^*\gamma^*$.  Rather
than use the formulae of the equivalent photon approximation,
which overestimates the $\gamma\gamma$ luminosity by about 15\%
in the $W$ range reported in this article, we rely
on a Monte Carlo generator based on the formalism of
Budnev~{\em et~al.}\cite{ref:Budnev} Only transverse photon
polarization states are considered.
The hadron component of each nearly real photon is modeled by a
$\rho$-pole form factor.  The small but finite $q^2$ values of each
photon cause these form factors to suppress the $\gamma\gamma$
luminosity by approximately 30\% for the kinematic range explored by
this experiment.

The event generation is completed by
incorporating the Brodsky and Lepage\cite{ref:Brodsky} QCD model
for $\gamma\gamma\rightarrow M^+M^-$; but
for simplicity, the explicit $\sin^{-4}\theta^*$ dependence of
Eq.~(\ref{eq:Brodsky}) is replaced with a flat $\cos\theta^*$ dependence.
The total angular integral over $\cos\theta^*$ is scaled to the
appropriate prediction of Eq.~(\ref{eq:Brodsky}).
The event generation is carried out separately for several ranges of
$W$; each sample represents in excess of 2.9~$\rm fb^{-1}$ of
integrated luminosity.

The rest of the Monte Carlo simulation involves tracking the
generated particles through a detector simulation
package based on GEANT\cite{ref:GEANT}.
In an effort to simulate noise in the detector elements,
random events selected by a dedicated trigger are embedded
into the Monte Carlo sample.  After a simulation of the trigger
logic is performed, the events are then reconstructed in
exactly the same manner as for data.  The accuracy of most of the
components of the
event simulation is demonstrated in Sec.~\ref{sec:calibration}
by the comparison with two calculable QED processes.

\subsection{\label{sec:selection} Event Selection}
The $e^+e^-M^+M^-$ final state system arises
from the collision of quasi-real photons
radiated by the electrons.  Since the electrons typically scatter at
very small angles, they generally go unmeasured in the experiment.
Likewise, the $\gamma\gamma$ collision axis is nearly coincident with
the electron beam axis, leaving the event $p_\perp$ unchanged.
The energies of the radiated photons are not equal in general,
so typically the final state system is
Lorentz boosted with respect to the laboratory.

The event selection cuts employed exploit these general
features of two-photon scattering at $e^+e^-$ colliders to suppress
the large background of annihilation events.  Exactly two oppositely
charged particles consistent with originating from the interaction point
(within 5~mm transverse to the beam direction and 5~cm along it) are required
to leave tracks in the drift chambers.
The scalar sum of the two track momenta must be less than 8~GeV/$c$
to suppress QED events, and
the total amount of energy deposited in the electromagnetic
calorimeter must be less than 8~GeV (the beam energy is
approximately 5.3~GeV).
A total of 6.3 million triggered events survive these preselection cuts.

As two-photon events tend to be well balanced in $p_\perp$,
a cut is imposed to require that the total
charged transverse momentum of the event be less than 200~MeV/$c$.
Annihilation produced tau events ($e^+e^-\rightarrow \tau^+\tau^-$),
one of the principal backgrounds to this measurement, tend
not to satisfy this requirement owing to the missing momentum
carried away by the neutrinos.  (The expected level of tau pair contamination
is addressed in Sec.~\ref{sec:bknd}.)
A cut is similarly imposed to require that the
two tracks be opposite in $\phi$ to within 50~mrad.
To further ensure that there are no undetected particles other than
the scattered electrons,
the polar angle of the missing momentum vector of the event must
point to within $10^\circ$ of the beam axis.
Cosmic rays often trigger the experiment and may be classified as
two-photon events if they pass near the interaction point.
To reduce their contamination, as well as the
contamination from annihilation events, the two tracks
must pass an acolinearity requirement:  the angle between the two
tracks must be more than 100~mrad away from back-to-back.
Finally, the two tracks are required to project to well understood
regions of the detector and must fire a two-track trigger.
The angular cut is $|\cos\theta|<0.7$ for each track. This directly
restricts the angular acceptance in the $\gamma\gamma$ center-of-mass
frame, and a cut is made to select events with $|\cos\theta^*|<0.6$
since there is little acceptance outside this range.
With all event shape
requirements applied, the data sample is reduced to 488,000 events.

Events containing electrons are suppressed by cutting on the quantity
$E_{\rm CC}/p$.  Since
electrons will deposit most of their energy in the CsI calorimeter,
the ratio of the energy deposit $E_{\rm CC}$ to the magnitude of the momentum
$p$ will peak at 1.0.  A cut is made to select only events with tracks
satisfying $E_{\rm CC}/p<0.7$. Events
with neutral energy observed in the detector are also rejected by
requiring that the total amount of energy in the CsI
calorimeter unassociated with tracks, $E_{\rm neut}$, be less than 500~MeV.
Both barrel and
endcap calorimeter energy clusters are considered.  Events with
explicitly reconstructed $\pi^0$s in them---identified
as having two unmatched, isolated
showers of at least 50~MeV in energy that combine to within 15~MeV/$c^2$ of the
nominal $\pi^0$ mass---are also rejected.  This leaves 172,000 events
satisfying all cuts but the muon rejection criteria to be discussed
in Sec.~\ref{sec:rejection}.  Table~\ref{tab:evtshape} summarizes all
cuts used in the two-prong selection.

Charged pions and kaons can be separated by $dE/dx$ or
time-of-flight measurements.  Unfortunately, the flight times for
pions and kaons merge to within experimental uncertainty for momenta
greater than about 1.2~GeV/$c$ using the CLEO~II time-of-flight system.
Furthermore,
the $dE/dx$ energy loss bands using the CLEO~II drift chamber
cross at about 1.0~GeV/$c$, and are only
separable at the  $2\sigma$ level for momenta larger than about
2.0~GeV/$c$.  Because of these limitations, pions and kaons will not be
separated, and a combined cross section will be
reported.  Thus, the ratio of the $K^+K^-$ to $\pi^+\pi^-$ cross
sections cannot be measured, and this forfeits one test of the
$\gamma\gamma\rightarrow M^+M^-$ models.
The two-photon production of proton-antiproton pairs is negligible
owing to the steep $W$ dependence of the cross
section\cite{ref:CLEOppbar,ref:Farrar}.

\subsection{\label{sec:rejection} Muon Rejection}
The primary challenge of the measurement of the two-photon production
of charged pion and kaon pairs is to reduce the enormous two-photon background
of muon pairs. Equation~(\ref{eq:Brodsky}) predicts that the meson pair
cross section will have an extra $W^{-4}$ dependence
relative to the $\mu^+\mu^-$
cross section, becoming several orders of magnitude smaller for $W$
of a few GeV/$c^2$.

Muons can be identified by their
penetration to a set of tracking chambers located outside several
interaction lengths of material.  This can be accomplished with the
CLEO~II muon chamber system;
however, muons must have a momentum $p \gtrsim 1$~GeV/$c$ in order
to penetrate the first layer of iron and
produce hits in at least the first set of tracking layers.
This restricts $W$ to be larger than approximately 2~GeV/$c^2$ for adequate
muon rejection.
In addition, the acceptance loss in
requiring both tracks to project sufficiently far from the cracks
in the muon chamber system to
limit the uncertainty due to multiple scattering is approximately 50\%.

An alternative method of separating muons from hadrons
capitalizes on the different energy deposit patterns within the CsI
calorimeter.  It has the advantage of full azimuthal coverage (no
cracks) and lower momentum reach, but it must be accomplished using
the limited depth of the electromagnetic calorimeter.
Muons typically leave a narrow trail of ionization and lose little
energy as they traverse the 30~cm depth of the calorimeter.
Hadrons, too, generally pass right through since the electromagnetic
calorimeter is only about an interaction length deep.  Occasionally,
though, a pion or kaon will suffer an inelastic collision and deposit
a significant amount of energy over a large lateral area.
The energy deposited in the CsI calorimeter and the lateral shower size
can be used to tag one of the mesons within two-prong
events and thus suppress the muon pair background.

The lateral shower size is computed
by calculating the root-mean-square (RMS) width of the shower
relative to the shower
center-of-gravity.  The center-of-gravity is computed by finding the
energy-weighted mean of the locations of each crystal within all
matched energy clusters, where the matching algorithm includes all showers
within about 30~cm of the track projection.
The shower width is calculated by taking the energy-weighted
root-mean-square of the distance each crystal lies away from the
center-of-gravity, which ought to be small for muons and
non-interacting hadrons.  The situation is complicated by the track
curvature in the plane transverse to the beam direction,
which makes the width depend
on the transverse momentum. To separate this effect, the width is
calculated in the $r$--$z$ plane only as Fig.~\ref{fig:shape_diagram}
shows, where the $z$-axis is aligned with the beam direction and $r$
points from the interaction point to energy cluster.
The RMS width
is calculated along just the $z$-axis, and is
multiplied by $\sin\theta_{\rm cog}$:
\begin{eqnarray}
R_{\rm z} & = & \sqrt{{\sum_i E_{\rm i} d_{\rm i}^2}\over{\sum_i E_{\rm i}}} \\
d_{\rm i} & = & (z_{\rm i} - z_{\rm cog})\sin\theta_{\rm cog} \nonumber
\end{eqnarray}

The discrimination power of the shower energy and lateral size is evident in
Figs.~\ref{fig:mupi_shape}a and \ref{fig:mupi_shape}b.  These
two-dimensional plots show the total matched energy $E_{\rm CC}$
along the ordinate
and the shower width $R_{\rm z}$ along the abscissa for particle
momenta between 2 and 3~GeV/$c$.  Tracks are restricted to the barrel
of the detector ($|\cos\theta|<0.7$).
Muons selected from  $e^+e^-\rightarrow \mu^+\mu^-\gamma$
are shown in (a), and
charged pions from $\rho^\pm$ decays in
two-prong annihilation produced tau events
are shown in (b).
A two-dimensional cut is very effective at separating muons from
interacting charged hadrons. The lines shown in Fig.~\ref{fig:mupi_shape}
define the acceptance region as $E_{\rm CC}>400$~MeV and $R_{\rm z}>3.5$~cm.
For momenta less than 3~GeV/$c$, only about 1 in 700 muons pass both
cuts.

Figure~\ref{fig:pi_shape} shows the
efficiency for charged pions to lie in this
acceptance region as a function of momentum ($|\cos\theta|<0.7$).
Data from $\rho^\pm$ decays are shown by the solid circles, and the
efficiency is consistent with that determined from pions in $K_{\rm s}^0$
decays as well as pions in four-prong events.  A Monte Carlo
simulation based on the
detector simulation
package (histogram) shows good agreement with the data.
The charged kaon showering efficiency is expected to be similar to
that of charged pions based on a Monte Carlo simulation.

The most efficient use of the shower selection criteria is to require
just one track in a two-prong event to satisfy the shower shape cuts.
This yields a
75\% efficiency at large momenta, and a rejection factor for muon pair
events of approximately $350$.  This allows a reach in $W$ of up to
5.1~GeV/$c^2$ before the two-photon muon production cross section
equals that predicted for charged pion and kaon pairs combined.
An advantage of following this procedure
is that it provides a self-checking method of
examining the purity of the data sample.  One can tag one charged meson as
passing the shower shape cuts and then measure the shower shape
efficiency of the other particle.  This method would also be
sensitive to any deviation of the charged kaon efficiency from that
of charged pions.
The empty circles in Fig.~\ref{fig:pi_shape} represent this self-measured
efficiency for two-prong events satisfying the two-photon event
selection cuts summarized in Table~\ref{tab:evtshape}. The good agreement with
the $\rho^\pm$ data for
momenta less than 2~GeV/$c$ suggests that there is little muon
contamination in that region.
For momenta above 2~GeV/$c$, however,
the self-measured efficiency is significantly lower than expected.
The inferred contamination is well described by known background
sources of muons (see Sec.~\ref{sec:bknd}), and can easily be reduced to
negligible levels using the muon chamber system since it is fully efficient
at identifying muons (and thus capable of suppressing them)
at these momenta.
The tight geometric acceptance to avoid the cracks
in the muon system is not necessary to accomplish this
final level of filtering.
Therefore, to reject muons, we demand that at least one track
satisfies the shower shape criteria, and that no muon chamber hits
correlate to either track.

\section{\label{sec:results} Results}

\subsection{\label{sec:calibration} Physics Calibration}
The cross section measurement of a reaction
similar to the two-photon production of charged pion and kaon pairs, but
theoretically well understood,
checks our understanding of the detector
efficiencies as well as the modeling used at the generator level.
Such a process is the two-photon production of muon pairs
($e^+e^-\rightarrow e^+e^-\mu^+\mu^-$).  These events have a
topology similar to $e^+e^-\rightarrow e^+e^-M^+M^-$,
and the process can be calculated to high
precision within QED.  However, the two-photon event shape cuts
described in Sec.~\ref{sec:selection} select not only two-photon events,
but also radiative annihilation events.  In particular, the process
$e^+e^-\rightarrow \mu^+\mu^-\gamma$ cannot be distinguished from
$e^+e^-\rightarrow e^+e^-\mu^+\mu^-$ when the radiated photon
is at a small angle with respect to the beam axis---aside from the
typically larger invariant mass of the two muons---since the $\mu^+\mu^-$
center-of-mass will be Lorentz boosted as in two-photon collisions.

All of the two-prong selection cuts discussed in Sec.~\ref{sec:selection}
are used to
select the muon pair events.  In addition, to obtain a clean sample of muons,
both tracks are required to correlate to hits in the innermost superlayer
of the muon
chamber system.
The invariant mass distribution of the selected
events from 500~$\rm pb^{-1}$ of data is shown in
Fig.~\ref{fig:mumu} (solid points).
The spectrum does not begin until about 2~GeV/$c^2$ as the
momentum of each particle must be large enough to penetrate the first
layer of iron.
The high $W$ cutoff is caused by the cut on the scalar sum of
the track momenta.  The histograms shown in Fig.~\ref{fig:mumu}
represent a Monte Carlo simulation of the reactions
$e^+e^-\rightarrow e^+e^-\mu^+\mu^-$ (solid) and
$e^+e^-\rightarrow \mu^+\mu^-\gamma(\gamma)$ (dashed).  The two-photon
generator is the SIXDIA generator taken from Vermaseren\cite{ref:SIXDIA}.
The $e^+e^-$ annihilation generator is the FPAIR generator by Kleiss
and van~der~Mark\cite{ref:FPAIR}.
The two-photon events peak at low $W$, while the
annihilation events peak at the machine center-of-mass energy and
have a long radiative tail toward smaller values of $W$.
A 3\% downward correction has been applied to the
Monte Carlo trigger efficiency to account for differences with the
efficiency measured from data.  This correction was determined using
processes other than $e^+e^-\rightarrow e^+e^-\mu^+\mu^-$ (see
Sec.~\ref{sec:trigger}) and carries a 3\% systematic uncertainty.
The uncertainty and magnitude of this correction is smaller than that
for charged pion and kaon pairs since the interaction of muons in the
calorimeter is better understood.

As can be seen, the resulting simulation does well at
reproducing the event yield across all of $W$.
The overall areas of
the two distributions agree to within 3\%, with the Monte Carlo
slightly underestimating the event yield.
This suggests
that the modeling of the detector efficiencies (including the trigger
efficiency) are well understood
and that the theoretical calculations of Refs.~\ref{ref:SIXDIA} and
\ref{ref:FPAIR} are in agreement with the data.

\subsection{\label{sec:bknd} Backgrounds}

Despite the rejection power of the event selection cuts described in
Sec.~\ref{sec:selection}, some background events still manage to contaminate
the charged pion and kaon pair data sample.
The two-photon background
$e^+e^-\rightarrow e^+e^-\mu^+\mu^-$ is
only expected to contribute about 1~event for $W>4$~GeV/$c^2$
according to Monte Carlo calculations when one track is required to
satisfy the shower tag described in Sec.~\ref{sec:rejection}.
A larger contribution comes from the tail of
radiative muon pair events ($e^+e^-\rightarrow \mu^+\mu^-\gamma$), where
$30\pm10$ events are expected to pass the two-photon event shape
requirements and the shower tag requirement.
The uncertainty arises from
the uncertainty in the muon rejection factor.
This contamination was noticed in the
shower tag efficiency extracted from the charged pion and kaon
pair data sample itself
(shown as empty circles in Fig.~\ref{fig:pi_shape}),
where $p>2$~GeV/$c$ roughly corresponds to $W>4$~GeV/$c^2$;
but it is reduced to negligible levels when a
cut is imposed to exclude any events with muon chamber hits correlating
to either track.

Significant non-muon
backgrounds arise from the tau pair production of charged mesons. In
particular, the small sample of $e^+e^-\rightarrow\tau^+\tau^-\rightarrow
\pi^+\pi^-\nu_\tau\overline{\nu}_\tau$
($\pi$--vs--$\pi$) events that pass
the two-photon
event shape cuts described in Sec.~\ref{sec:selection} are problematic.
{}From a Monte
Carlo simulation using Standard Model parameters for the $\tau$ decays,
$11\pm2$ $\pi$--vs--$\pi$ events are expected to pass the shower
tag on one of the tracks.
An additional $4\pm3$ events are expected to pass from the
$K$--vs--$\pi$, $K^*$--vs--$\pi$, and $\rho$--vs--$\pi$ channels.
The $\mu$--vs--$\pi$ background is negligible when tracks are
required not to correlate to muon chamber hits.
These events are spread more or less uniformly across the invariant mass
spectrum from about 1~GeV/$c^2$ to the cut at 8~GeV/$c^2$
(the invariant mass is calculated from the two charged final state
particles assuming pion masses);
therefore, we expect about 2 tau events for each 1~GeV/$c^2$ wide bin in
$W_{\pi\pi}$.  This residual background becomes more than 5\% of the
total two-prong data sample for $W_{\pi\pi}>3$~GeV/$c^2$, but it
is subtracted from the
two-prong data sample in the calculation of the two-photon charged
pion and kaon pair cross section

\subsection{\label{sec:cross_section} Charged Pion and Kaon Pair Cross Section}

The extraction of the two-photon cross section from data involves
comparing the two-prong event yields from data and Monte Carlo after
all selection cuts have been imposed, and then scaling the average Monte
Carlo $\gamma\gamma$ cross section appropriately:
\begin{equation}
\sigma_{\rm meas}\, (\gamma\gamma\rightarrow \pi^+\pi^- + K^+K^-)\; =\;
{N_{\rm data}\over N_{\rm MC}}{L_{\rm MC}\over L_{\rm data}}
\;\overline{\sigma}_{\rm MC}\, (\gamma\gamma\rightarrow \pi^+\pi^- + K^+K^-)
\label{eq:sigmaextract}
\end{equation}
A 5\% downward correction to the Monte Carlo trigger efficiency
(in contrast to a 3\% correction to the muon pair trigger efficiency)
is applied to reflect measurements made from data.
The factors $L_{\rm MC}$ and $L_{\rm data}$ are the integrated
luminosities of the Monte Carlo and data
samples respectively, and the number of events $N_{\rm data}$ and
$N_{\rm MC}$ are computed bin by bin in $\cos\theta^*$ and
$W_{\pi\pi}$ (the two-prong invariant mass calculated assuming pion
masses for the particles).
The Monte Carlo cross section $\sigma_{\rm MC}$,
integrated over the angular range $|\cos\theta^*|<0.6$, is given by:
\begin{equation}
\sigma_{\rm MC}(W_{\pi\pi})\; = \;
{ {135~{\rm nb}} \over {W^6_{\pi\pi}} } \;
\bigg[1 + 2.2\bigg( { {W_{\pi\pi}} \over {W}_{\rm KK}}
\bigg)^6\;\bigg]    \label{eq:sigma_QCD}
\end{equation}
The cross section
$\sigma_{\rm MC}$ represents the combined cross
section for the production of $\pi^+\pi^-$ and $K^+K^-$.  The factor
2.2 comes from $(f_{\rm K}/f_\pi)^4$ in the QCD model of Brodsky and Lepage,
and the factor $(W_{\pi\pi}/W_{\rm KK})^6$ comes from
correcting $W$ to reflect kaon masses.  The average value of this
cross section for each bin in $\cos\theta^*$ and $W_{\pi\pi}$ is used
in Eq.~(\ref{eq:sigmaextract}), but the dependence of $\sigma_{\rm meas}$
on our particular choice of $\sigma_{\rm MC}$ is minimal since it
cancels in Eq.~(\ref{eq:sigmaextract}) to first order.

The data used in the extraction of the cross section
are the two-photon two-prong candidates that
pass the shower shape cut on at least one track and that exhibit
no muon chamber tracks.  A total of 6286 events satisfy these cuts.
The final result, shown along with the QCD models of
Brodsky and Lepage\cite{ref:Brodsky}, and Benayoun and
Chernyak\cite{ref:Benayoun}, is presented
in Fig.~\ref{fig:pipi} as a function of $W_{\pi\pi}$.  Only the statistical
errors are shown.  The yield in each $W_{\pi\pi}$ bin is given in
Table~\ref{tab:events} along with the detection efficiencies
and measured cross sections.
The ratio of the $\gamma\gamma\rightarrow K^+K^-$
cross section to the $\gamma\gamma\rightarrow\pi^+\pi^-$ cross
section in the Benayoun and Chernyak model is 1.08 rather than 2.20
as in the Brodsky and Lepage model, but Fig.~\ref{fig:pipi}
demonstrates that the sum of the two cross sections is nearly
the same in both models.

The measured cross section agrees well in shape with
the QCD models over the entire measured range in $W_{\pi\pi}$, though
the normalization of the models is approximately 40\% below the data.
There is no compelling reason to
expect that the cross section should exhibit good agreement at
low $W_{\pi\pi}$, however, since perturbation theory is not expected
to be within kinematic range
and resonance formation should contribute.
The last point plotted in Fig.~\ref{fig:pipi} represents
an upper limit at the 90\% confidence level based on the observation
of zero events (with a background expectation of two) in the $W_{\pi\pi}$
range from 4 to 5~GeV/$c^2$.
When the veto on events containing reconstructed $\pi^0$s is
relaxed, two events do fall into this bin; but they are consistent
with the tau process $e^+e^-\rightarrow\tau^+\tau^-\rightarrow
\rho^+\pi^-\nu_\tau\overline{\nu}_\tau$, in agreement with
the expected level of tau background.

The evolution of the $\cos\theta^*$ distribution as a function of the
invariant mass $W$ can be used to
test the $\sin^{-4}\theta^*$ dependence of the QCD prediction.
The Mark~II measurement\cite{ref:MarkII} did not have sufficient
statistics to demonstrate such an angular distribution.
We report a measurement of the differential pion and kaon pair
cross section $d\sigma/d|\cos\theta^*|$ for four $W_{\pi\pi}$ bins,
each containing three $|\cos\theta^*|$ bins.
The result is shown in Figs.~\ref{fig:sigma_cos}a--d.  The angular
bins are equally spaced in $|\cos\theta^*|$ from 0.0 to 0.6, and the
$W_{\pi\pi}$ ranges are: 1.5--2.0, 2.0--2.5, 2.5--3.0, and
3.0--4.0~GeV/$c^2$.  The leading order QCD prediction for
$d\sigma/d|\cos\theta^*|$, taken from the Brodsky and Lepage
model\cite{ref:Brodsky}, is represented by the lines
drawn in each figure.  The measured cross
section does exhibit a rise
at small scattering angles consistent with that expected by leading order
perturbative QCD in the highest two $W_{\pi\pi}$ ranges.
The behavior in the first range ($1.5<W_{\pi\pi}<2.0$~GeV/$c^2$)
is opposite what one expects based on QCD, while that in the second region
($2.0<W_{\pi\pi}<2.5$~GeV/$c^2$) is flat in $\cos\theta^*$.
The disparity of the measured low $W_{\pi\pi}$ angular distribution with
leading order QCD may be related to the
non-perturbative nature of the cross section near the resonance
region, and in particular, to
the interference of the $f_2$ resonance with the
$\pi^+\pi^-$ continuum. Indeed, the angular dependence reported
here is similar to the angular dependence of the charged pion pair cross
section in the region near the $f_2$ (see for example J.~Boyer {\em
et al.}\cite{ref:f_two}, and references therein).

The Mark~II experiment performed a similar combined measurement of the
charged pion and kaon
pair cross section at the PEP $e^+e^-$ storage ring\cite{ref:MarkII}.
A direct comparison of our measured cross section with theirs is not
straightforward, however, because of differing theoretical treatments in the
extraction of the two-photon cross section.  The primary differences
involve the use of the equivalent photon approximation
in the Mark~II analysis, and the application of a $\rho$ form
factor to model the $q^2$ dependence of the virtual photons in our analysis.
The combined effect of these two factors leads to an approximate 45\% downward
shift of the Mark~II two-photon cross section relative to ours.  Further
differences include a smaller angular acceptance in the Mark~II measurement
($|\cos\theta^*|<0.5$) and a slightly different choice of $f_{\rm K}/f_\pi$.
When both analyses are treated in a similar manner,
the measured cross sections of CLEO and Mark~II
are compatible to within 10\%.

The TPC/Two-Gamma
Collaboration finds\cite{ref:TPCpipi} that the $\pi^+\pi^-$
cross section is about three times larger than the leading order
QCD prediction (as given by the Brodsky and Lepage model)
in the $W$ range from 1.5 to 2.5~GeV/$c^2$, and in
the smaller angular region $|\cos\theta^*|<0.3$.  Their kaon pair
production cross section, though, is more in line with QCD
in the measured $W$ range from 1.5 to 3.5~GeV/$c^2$,
with $|\cos\theta^*|<0.6$.
We find that in the narrow angular region
$|\cos\theta^*|<0.3$, the TPC/Two-Gamma data lie approximately 30\%
below our data on the combined cross section for pion and kaon pairs
in the $W_{\pi\pi}$ range from 1.5 to 2.0~GeV/$c^2$.  In this region
our data also are several times larger than the leading order QCD
prediction, see Fig.~\ref{fig:sigma_cos}.

\subsection{\label{sec:systematics} Systematic Effects}

Systematic uncertainties on the measured cross section
involve uncertainties in the trigger efficiency ($\sim5\%$)
and the shower shape efficiency.  The latter efficiency has been presented in
Fig.~\ref{fig:pi_shape} using various techniques, all in good
agreement with the detector simulation.
The stability of the measured cross section to the shower shape
criteria has been studied by varying the cuts on the particle energy
deposition in the calorimeter and the associated shower width by 25\%
of the nominal values.  This range of values changes the overall event
yield by as much as 50\% for small $W_{\pi\pi}$, but leaves the
extracted cross section unchanged to within 5\%.
Moreover, the cross section has been measured with the shower shape criteria
applied to both detected particles rather than just one.  Although the event
yield is reduced approximately by two-thirds,  the extracted cross section
deviates by less than 10\%.  The shower shape criteria have also been applied
to select charged pions in tau $\mu$--vs--$\pi$ events,
where several of the event shape cuts have been modified to select
events with missing momentum rather than suppress them.  The muon is
identified by its penetration to the muon chamber system as was done
for the calibration study in Sec.~\ref{sec:calibration}.  The
resulting cuts are loose enough to select additional tau
channels---notably channels containing neutral energy---but when all
tau channels are included into the Monte Carlo generator, the
resulting event yield agrees with that from data to 5\%.
Finally, the two-photon cross section has been measured using only the muon
chamber system in the limited kinematic region of adequate muon pair
suppression.  Results of this alternate approach confirm the functional
dependence of our measurement of the cross section in both $W_{\pi\pi}$ and
$|\cos\theta^*|$, but disagree somewhat on absolute normalization.  Based on
this study, and the previous tests of the shower shape efficiency, we assign a
20\% systematic error to the overall normalization of our measurement, and a
10\% systematic error on the point-to-point variation.

In addition to the modeling uncertainties described above,
the measured cross section
has a slight systematic dependence on the particular choice for the ratio
of the kaon pair cross section to the pion pair cross section in the
Monte Carlo generator through Eq.~(\ref{eq:sigmaextract}).  This ratio is
taken to be 2.2, but if it is reduced to unity as motivated by Benayoun
and Chernyak\cite{ref:Benayoun}, the reported cross section will also
be reduced.  The reduction is largest at small $W_{\pi\pi}$, where the cross
section is reduced by 12\% for $W_{\pi\pi}<1.8$~GeV/$c^2$. For $W_{\pi\pi}$
in the range from 1.8 to 2.75~GeV/$c^2$, the reduction is approximately
8\%; and for $W_{\pi\pi}$
larger than 2.75~GeV/$c^2$, the reduction is 3\% or less.
Finally, as noted before, removing the $\rho$ form factors in the
modeling of the $q^2$ dependence of the virtual photons reduces our
measured cross section by approximately 30\%.

\section{\label{sec:conclusion} Conclusion}

We have presented data on the combined two-photon production of
charged pion and kaon pairs from a sample of 1.2~$\rm fb^{-1}$
collected at CESR.
The $W_{\pi\pi}$ dependence of the measured cross section
in the angular region $|\cos\theta^*|<0.6$ shows
reasonable agreement with leading order perturbative QCD predictions across
the entire measured range from 1.5 to 5.0~GeV/$c^2$, though the QCD based
models lie approximately 40\% below the data.  The $\cos\theta^*$
dependence of the cross section exhibits a rise at small scattering angles
consistent with that expected by QCD for $W_{\pi\pi}>2.5$~GeV/$c^2$.
The magnitude of the measured cross section is in good agreement with
results from the Mark~II experiment, but is in disagreement
by about 30\% with results from the TPC/Two-Gamma experiment.
Systematic uncertainties on the extracted cross section from detector
acceptance effects amount to 20\%, while different theoretical treatments in
the extraction of the cross section lead to variations of up to 30\%.
The reported measurement extends the reach in $W$
further than previous experiments, allowing the observation of a
transition in the cross section from non-perturbative to
perturbative behavior.
Sufficient rejection of leptonic events is achieved through the novel
use of the electromagnetic calorimeter as a tag on early interacting
hadrons, which removes some of the fiducial and momentum constraints
of the conventional muon chamber system.

\centerline{\bf ACKNOWLEDGEMENTS}
\smallskip
We gratefully acknowledge the effort of the CESR staff in providing us with
excellent luminosity and running conditions.
J.P.A. and P.S.D. thank
the PYI program of the NSF, I.P.J.S. thanks the YI program of the NSF,
G.E. thanks the Heisenberg Foundation,
K.K.G., I.P.J.S., and T.S. thank the
%
%
 TNRLC,
K.K.G., H.N.N., J.D.R., T.S.  and H.Y. thank the
OJI program of DOE
and P.R. thanks the A.P. Sloan Foundation for
support.
This work was supported by the National Science Foundation and the
U.S. Dept. of Energy.

%

%
\begin{figure}[p]
\centerline{\epsfbox{shapecut.ps}}
\caption{\label{fig:shape_diagram}Diagram illustrating the shower width
calculation. The distance $d_{\rm i}$ is taken to be the distance
along $z$ from crystal~$i$ to the shower center-of-gravity, multiplied
by $\sin\theta_{\rm cog}$.}
\end{figure}
%
%
\begin{figure}[p]
\centerline{\epsfbox{mupi_shape.ps}}
\caption{\label{fig:mupi_shape} Scatter plots of the matched
CsI energy versus the shower width for muons~(a) and pions~(b).  A
particle is tagged as a hadron if $E_{\rm CC}>400$~MeV and $R_{\rm z}>3.5$~cm.}
\end{figure}
\newpage
\begin{figure}[p]
\centerline{\epsfbox{shwr_effic.ps}}
\caption{\label{fig:pi_shape} The efficiency for charged pions and kaons
to satisfy the shower tag as a function of momentum.  Data shown
include pions from $\rho^\pm$ decays as solid circles ($1$--vs--$\rho$)
as well as pions
and kaons from two-prong events as empty circles ($M^+M^-$).  A Monte Carlo
simulation of the charged pion efficiency is shown by the histogram.}
\end{figure}
%
\begin{figure}[p]
\centerline{\epsfbox{mumu_new.ps}}
\caption{\label{fig:mumu}
Invariant mass distribution of muon pair
events passing two-photon event selection criteria.  The solid histogram
represents a simulation of $e^+e^-\rightarrow e^+e^-\mu^+\mu^-$, and
the dashed histogram a simulation of $e^+e^-\rightarrow \mu^+\mu^-\gamma$.}
\end{figure}
\newpage
\begin{figure}[p]
\centerline{\epsfbox{pipi_new.ps}}
\caption{\label{fig:pipi}
Measured cross section for the two-photon production of charged pion
and kaon pairs as
a function of $W_{\pi\pi}$ in the angular region $|\cos\theta^*|<0.6$.
Only statistical errors are shown. The
leading order QCD predictions by Brodsky and Lepage, and Benayoun and Chernyak,
are shown by the solid and dashed curves respectively.}
\end{figure}
\newpage
\begin{figure}[p]
\centerline{\epsfbox{sigma_cos_new.ps}}
\caption{\label{fig:sigma_cos}
The $\cos\theta^*$ dependence of the charged pion and kaon pair cross section
for four
bins in $W_{\pi\pi}$: (a) 1.5--2.0~GeV/$c^2$, (b) 2.0--2.5~GeV/$c^2$,
(c) 2.5--3.0~GeV/$c^2$,
and (d) 3.0--4.0~GeV/$c^2$.  The leading order QCD prediction by Brodsky and
Lepage is represented by the solid curve.}
\end{figure}
\newpage
\begin{table}[p]
\begin{tabular}{rll}
system      & & efficiency (\%)\\ \hline
time-of-flight    &  & $98.0\pm1.0$   \\
calorimeter       &  & $92.0\pm4.0$   \\
tracking          & early running & $88.0\pm2.0$ \\
                  & late running  & $95.0\pm1.5$ \\
\end{tabular}
\caption{\label{tab:trigger} The measured efficiencies of various
components of the two-prong trigger.}
\end{table}
%
%
%
\begin{table}[p]
    \begin{tabular}{rll}
      {\bf 1.} & $N_{\rm track} = 2$ & (exactly two charged tracks) \\
      {\bf 2.} & $ \sum Q_{\rm i} = 0$ & (net charge of zero) \\
      {\bf 3.} & $d_\perp (i)<$ 5 mm (i=1,2)&(impact parameter $\perp$ beam)\\
      {\bf 4.} & $d_\parallel (i) <$ 5 cm (i=1,2)&(impact parameter
                                                  $\parallel$ beam)\\
      {\bf 5.} & $\sum |p|<$ 8 GeV/$c$ & (low total charged momentum) \\
      {\bf 6.} & $\sum E_{\rm CC} <$ 8 GeV & (low total calorimeter energy) \\
      {\bf 7.} &  $p_\perp({\rm event})\equiv|\vec{p}_{1\perp} +
                   \vec{p}_{2\perp}|<200$~MeV & ($\perp$ momentum balance) \\
      {\bf 8.} &  acoplanarity$\equiv|\pi - \Delta\phi|<0.05$
               & (back-to-back in $\phi$) \\
      {\bf 9.} & acolinearity$\equiv|\pi - \arccos\hat{p}_1\cdot\hat{p}_2|
                 >0.1$ & (not back-to-back in 3D) \\
      {\bf 10.} & $\theta_{\rm miss} > 170^\circ$ or $<10^\circ$
                & (missing momentum $\parallel$ beam) \\
      {\bf 11.} & $|\cos\theta_{\rm i}|<0.7$ (i=1,2) &
                                                (tracks project to barrel) \\
      {\bf 12.} & $|\cos\theta^*|<0.6$ & (large angle scattering in c.o.m.)\\
      {\bf 13.} & $E_{\rm CC}/p < 0.7$ & (electron suppression) \\
      {\bf 14.} & $E_{\rm neut} < 500$~MeV & (low unmatched energy) \\
      {\bf 15.} & $\pi^0$ veto & (no reconstructible $\pi^0$s) \\
      {\bf 16.} & Dedicated two-prong trigger & (trigger requirement) \\
      {\bf 17.} & Muon suppression  & (see
                  Sec.~\protect\ref{sec:rejection})  \\
    \end{tabular}
  \caption{\label{tab:evtshape} Event shape and acceptance cuts used in
           the two-prong selection.}
\end{table}
%
%
\begin{table}[p]
\begin{tabular}{cdrc}
$W_{\pi\pi}$ (GeV/$c^2$) & Efficiency (\%) & Events &
$\sigma(\gamma\gamma\rightarrow\pi^+\pi^- + K^+K^-)$ (nb) \\ \hline
$1.5-1.6$     & 2.6 & 245  & $20.5\pm1.8$ \\
$1.6-1.7$     & 3.1 & 193  & $13.1\pm1.2$ \\
$1.7-1.8$     & 3.6 & 161  & $11.5\pm1.2$ \\
$1.8-1.9$     & 4.0 & 108  & $6.44\pm0.95$ \\
$1.9-2.0$     & 4.4 & 117  & $7.79\pm1.08$ \\
$2.0-2.1$     & 4.8 & 84   & $5.13\pm0.72$ \\
$2.1-2.25$    & 5.4 & 54   & $3.18\pm0.53$ \\
$2.25-2.5$  & 6.2 & 54     & $2.08\pm0.35$ \\
$2.5-2.75$  & 7.3 & 23     & $1.03\pm0.23$ \\
$2.75-3.0$  & 8.3 & 16     & $0.75\pm0.22$ \\
$3.0-3.5$   & 9.9 & 21     & $0.59\pm0.15$ \\
$3.5-4.0$   & 12.0 & 4     & $0.052\pm0.048$ \\
$4.0-5.0$    & 15.0 & 0    & $0.041$ (u.l.)\\
\end{tabular}
\caption{\label{tab:events} The number of charged pion and kaon
pair candidates
passing all cuts for the $W_{\pi\pi}$ ranges shown in
Figs.~\protect\ref{fig:pipi}. Also
shown is the detection efficiency as determined by
Monte Carlo for $|\cos\theta^*|<0.6$, and the extracted cross section
for the process $\gamma\gamma\rightarrow\pi^+\pi^- + K^+K^-$. Only
statistical errors are shown.}
\end{table}
\end{document}